\documentclass[aps,amssymb,12pt,showpacs,showkeys,letterpaper]{revtex4}
\usepackage{graphicx}
\usepackage{amsmath}
\usepackage{epsfig}
\usepackage{bm}

\newcommand{\be}{\begin{equation}}
\newcommand{\ee}{\end{equation}}
\newcommand{\beq}{\begin{eqnarray}}
\newcommand{\eeq}{\end{eqnarray}}

\begin{document}
\title{Some features of a Schwarzchild Black Hole from a Snyder perspective }

\author{ Carlos Leiva }
 \email{cleivas62@gmail.com}
\affiliation{\it Departamento de F\'{\i}sica, Universidad de
Tarapac\'{a}, Casilla 7-D, Arica, Chile}

\date{\today}

\begin{abstract}

In this paper, the consequences of introducing a deformed Snyder-Kepler
potential   in the Schwarzchild metric are investigated. After this modification,
it is obtained a dynamically depending horizon with different
penetration radius for massive particles and light rays in radial
orbits. In the case of circular orbits, they all remain untouched.

\end{abstract}



\maketitle
\section{\label{sec:Int}Introduction}

The main stream research on very high energy physics and on possible
candidates for a suitable quantum gravity theory has lead to the
idea of the existence of a fundamental length. At energies
corresponding to this length it is expected to find  a totally new
non commutative physics that even could break the causal principle.
Among the main theories that back up this idea we can enumerate Loop
Quantum Gravity, String Theory and all their modifications,
proposals and variations, along with other efforts to give a
suitable quantum version for gravity. The minimal fundamental length
is usually identified with the minimal size of the elements of these
theories and the proposal usually  is that it approaches to the
order of Planck length.

We can considerate different ways to introduce a minimal fundamental
length, but many of them are incompatible with Lorentz symmetry that
is a corner stone in modern Physics,  and that is, of course,  a
very undesirable
consequence,\cite{MirKasimov:1996hv}\cite{Djemai:2003kd}\cite{Mirza:2002wm}.
In order to avoid that issue it is usual to use a proposal made by
H. Snyder, who in the $40's$ \cite{Snyder} postulated a modification
of the Heisenberg algebra that implies discrete spectra of the
spacetime operators. This modification can be seen as a one of the
$\kappa$-deformed spacetime modifications.

During decades the Snyder proposal was put aside due to the success
of the renormalization program in the standard model. However, the
emergence of candidates to quantum gravity that work on the base of
finite elements with minimal length has brought back the idea and
renewed interest on it.

One of the main challenges today, in this area, is to find ways to
experimentally test the discreteness of spacetime or the associated
non-commutativity of the canonical relations of the phase space
variables. There have been some efforts in this direction using
planetary data and different approaches from Newtonian Mechanic and
General Relativity dynamics , for instance Mignemi
\cite{Mignemi:2014jya} for the last case and  Romero
\cite{Romero:2003tu} and Leiva \cite{Leiva:2012az} for the former
case. Some features can be predicted from works like these,  but the
main general conclusion is that the new scenario must be proved
meanly in the presence of strong fields or in extraordinary
energetic situations. That¥s the reason to explore non commutativity
nearby black holes.

The paper is organized as follows: in the next section a short
review of the Snyder recipe is given and a modified Kepler potential
is recalled from \cite{Leiva:2012az}, in the third section
Schwarzchild metrics is modified and the main features are discussed
for massive particles and, finally, conclusions are given in the
last section.

\section{Snyder Space}

\subsection{Quantum Snyder Algebra}

 Quantum Snyder algebra is characterized by the following set of canonical
 parenthesis for the phase space variables:

\begin{equation}
[x_\mu,x_\nu]=ilM_{\mu \nu}, \label{nc}
\end{equation}
\begin{equation}
[x_\mu ,p_\nu]=i\delta_{\mu \nu}-ilp_\mu p_\nu, \label{nl}
\end{equation}
\begin{equation}
[p_\mu,p_\nu]=0.
\end{equation}

Where $\mu,\nu=0,1,2,3$.

 This has been the initial point for many
researches that have worked in many implications of the introduction
of a non commutative parameter $l$ since the Snyder's paper itself
\cite{Snyder}, others like \cite{ MirKasimov:1996hv},
\cite{Battisti:2010sr}, \cite{Leiva:2008at} and in recent days
\cite{Ivetic:2015cwa}.

\subsection{Classical Snyder Algebra}

Classical  $3$ dimensional  Euclidean Snyder Space is characterized
by its non linear commutation relations (now in the sense of Poisson
brackets), between the non commutative variables of the phase space
$\bar{x_{i}},\bar{p_{i}}$. A simple realization of these variables
in terms of standard commutative variables $x_{i},p_{i}$ can be done
after the recipe in \cite{Battisti:2010sr}:

\begin{eqnarray}
\bar{x}_{i}&=&x_{i}+l(xp)p_{i}, \\
\bar{p}_{i}&=&p_{i}
\end{eqnarray}

With $i,j=1,2,3$

This realization gives the following commutation relations:

\begin{eqnarray}
\{\bar{x}_i,\bar{x}_j\}&=& lL_{ij}, \\ \label{cnc}
\{\bar{x}_i,p_j\}&=&\delta_{ij}+lp_ip_j,\\ \label{cnl}
\{p_i,p_j\}&=&0,
\end{eqnarray}
where $l$ is as before, a parameter that measures  the deformation
introduced in the canonical Poisson brackets, and $L_{ij}$ is
defined as a dimensionless matrix proportional to the angular
momentum.

\subsection{Classical Snyder Newton  Potential}

 The Newtonian potential in terms of noncommutative variables is

\begin{equation}V=-\frac{MG}{\sqrt{\bar{x}^2}},
\end{equation}

  This can be implemented then in terms of the commutative space variables
 $x,p$ and at first order in $l$ :

\begin{equation}
V(x)=-\frac{\kappa}{\sqrt{x^2+2l(xp)^2}},
\end{equation}

so, using spherical coordinates:

\begin{eqnarray}
x&=&\rho \hat{\rho}, \\
p&=&m(\dot{\rho}\hat{\rho}+\rho
\dot{\theta}\hat{\theta}+\rho\dot{\varphi}\sin(\theta)\hat{\varphi},
\end{eqnarray}

and assuming $l m^{2} \dot{\rho}^2\ll 1$, the Snyder-Kepler
potential for a particle can be written as

\begin{equation}
V(\rho)=\frac{-MG}{\rho}(1-l m^{2} \dot{\rho}^2). \label{potential}
\end{equation}

\section{Snyder-Schwarzchild proposal }

Let's introduce this new potential instead of the classical Newton
potential in the Schwarzchild solution as a limit of weak
gravitation:

\begin{equation}
ds^2=-[1-\frac{2GM}{c^{2}\rho}(1-l m^{2} \dot{\rho}^2)]c^{2}dt^{2}
+[1-\frac{2GM}{c^{2}\rho}(1-l m^{2}
\dot{\rho}^2)]^{-1}d\rho^{2}+\rho^{2}d\Omega^{2}. \label{metric}
\end{equation}

From simple inspection it's easy to see that this proposal  is a
solution of vacuum Einstein field equation because the modification
depends just on $\dot{\rho}$. Indeed,  Ricci tensor and Ricci scalar
depend just on partial derivatives with respect to the variables
$\rho, \theta, \varphi$, so this proposal yields the Einstein
equation $G_{\mu\nu}=0$. On the other hand, the symmetries are still
the same, rotational and temporal displacement.

One of the main features that can be detected is related to horizon.
In standard Schwarzchild solution, does exist the well known radius
$\rho=\frac{2GM}{c^2}$ where the metric becomes singular and many
facts happen from the point of view of an exterior observer. This
horizon is modified in this formulation in the sense that the metric
is singular at $\rho=1-\frac{2MG}{c^2} (1-l m^2 \dot{\rho}^2)$.

The radius where metric is singular and where the light cones tilt
in that form that the inner sector is casually separated from the
outer sector of the black hole becomes dependent on the velocity and
mass of the particle.

Let's study this condition:

\begin{equation}
1-\frac{2MG}{\rho c^2} (1-l m^2 \dot{\rho}^2) > 0.
\end{equation}

This condition imposes a constrain on the velocity of the particle
that moves around the black hole:

\begin{equation}
\dot{\rho}^2> \frac{1}{l m^2} (1-\frac{c^2 \rho}{2MG}).
\label{condition}
\end{equation}

It is possible then study three cases:

\begin{enumerate}

\item
 $\rho>\frac{2MG}{c^2} $:

 In this case $1-\frac{\rho}{2MG}$ is always less than
 zero and there is no restrictions on the radial velocity.

 \item

 $\rho = \frac{2MG}{c^2}$:

 In this case \ref{condition} is always true and $g_{tt}=-l
 m^2\dot{\rho}^2$ and $g_{\rho \rho}=(l
 m^2\dot{\rho}^2)^{-1}$

 There is no horizon at $\rho=2MG$ because the metric is no singular
 at that point.

\item

$\rho<\frac{2MG}{c^2}$:

While condition \ref{condition} remains true, there is no horizon,
so it is possible to define a penetration radius:

\begin{equation}
\rho_p=\frac{2MG}{c^2}(1-l
 m^2\dot{\rho}^2).  \label{penet}
\end{equation}

 \end{enumerate}

 Due to the inclusion of the terms depending on the Snyder deformation, for an external observer
 it is possible to "see" a particle penetrating to the interior of the
 horizon of a black hole. The penetration radius is as little as the
 factor $l
 m^2\dot{\rho}^2$ that we supposed tiny ab initio, but can be
 significative for super massive particles with high speeds.

To see other effects,  and to explicit velocity, specially when the
particle approaches to $\rho=2MG$, let's write the massive particle
stipulation,  $-g_{\mu\nu} \frac{dx^\mu}{d\lambda}
\frac{dx^\nu}{d\lambda}=-1 $:

\begin{equation}
-[1-\frac{2GM}{c^{2}\rho}(1-l m^{2}
\dot{\rho}^2)]c^{2}\frac{dt^{2}}{d\lambda^2}
+[1-\frac{2GM}{c^{2}\rho}(1-l m^{2}
\dot{\rho}^2)]^{-1}\frac{d\rho^{2}}{d\lambda^2}+\rho^{2}\frac{d\Omega^{2}}{d\lambda^2}=-1.
\label{energy1}
\end{equation}

Now, despite the modification, we still have the Killing vectors
$A^\mu=(\partial_t)^\mu=(1,0,0,0)$ and
$B\mu=(\partial_\varphi)^\mu=(0,0,0,1)$, that using our deformed
metric lead to the conserved quantities:

\begin{eqnarray}
E&=&-[1-\frac{2GM}{\rho}(1-l m^{2} \dot{\rho}^2)]\frac{dt}{d\lambda}, \\
L&=& \rho^{2}\frac{d\varphi}{d\lambda}.
\end{eqnarray}

Where it has been chosen $c=1$ and $\theta=\frac{\pi}{2}$

Replacing the constants quantities $E$ and $L$ in \ref{energy1}:

\begin{equation}
-E^{2}+(\frac{d\rho}{d\lambda})^{2}+(1-\frac{2GM(1-lm^{2}\dot{\rho}^2)}{\rho})[\frac{L^{2}}{r^{2}}+1]=0.\label{energy2}
\end{equation}

The form of energy is no longer suitable of being divided into a
dynamical term (depending just on velocities) and an effective
potential (depending just on the coordinates), but it is noticeable
that circular orbits are untouched by this proposal because in that
case $\dot{\rho}=0$, the deformation becomes  zero and the standard
case is recovered.

On the other hand, even though the effective potential
$(1-\frac{2GM(1-lm^{2}\dot{\rho}^2)}{\rho})[\frac{L^{2}}{r^{2}}+1]$
contains terms that depend on radial velocity of the particle, its
shape  remains the same, just modulated by the modification
introduced. In fact, the radial velocity can be expressed as:

\begin{equation}
\dot{\rho}^2=
\frac{(\mathfrak{E}+\frac{GM}{\rho}-\frac{L^2}{2\rho^2}+\frac{GML^2}{\rho^3})}{(\frac{1}{2}+\frac{GM}{\rho}l
m^{2}+\frac{GML^2}{\rho^3}l m^{2})}.
\end{equation}

Where it has been identified $\frac{dt}{d\lambda}=\dot{\rho}$ and
defined $\mathfrak{E}=\frac{1}{2}(E^{2}-1)$

 It is possible to see that  the condition  $\dot{\rho^2}=0$ does not depend on the
Planck term.

 Obtaining $\dot{\rho^2}$ from \ref{energy2},  $L=0$ for radial
orbits and replacing in \ref{penet} (with $c=1$), we can see that
the penetration radius of a particle is:

\begin{equation}
\rho_p=2MG(1-E^2lm^2),
\end{equation}
that can be considerably less than $2MG$, depending on the energy of
the particle. So, for a enough energetic and massive particle an
exterior observer can "see" how it falls into the singularity.

Let's take a look on the penetration radius of a light ray in order
of having a notion of what is really possible to see from the
exterior. For that case we can replace $m \dot{\rho}$ and use
$p_\rho$ instead or, if we deal with radial orbits, just use $p$. It
is necessary also to let $d\tau=0$. So we have:

\begin{equation}
0=[1-\frac{2GM}{\rho}(1-l p)]dt^{2}
+[1-\frac{2GM}{\rho}(1-lp]^{-1}d\rho^{2}.
\end{equation}

Under the requirement that the penetration speed must be more than
zero, we have a penetration radius of a ray light:

\begin{equation}
\rho^{\textsl{light}}_p=2MG(1-lp^2).
\end{equation}
So, light rays can penetrate  (or escape), depending on its
momentum. Anyway, to hypothetically reach the singularity the wave
length should be of the order of the Planck scale, this is a
condition that matches the idea that at that scale Physics can be
totally different to the every day experience.

For the most part of high energy light rays it is possible to go
beyond the Schwarzchild radius and it could be imaginable to get
information from the interior of the black hole through a skin
penetration like phenomenon.

\section{Final Remarks}

It has been introduced a mass/velocity depending deformation to the
Schwarzchild metric that correspond to consider a minimal
fundamental length. Under this deformation the metric conserves the
symmetries and this fact is used to take a look on some variations
of the horizon of the black hole.  Effectively the introduction of
the Snyder factor changes kinematically the Schwarzchild radius and
allows particles and light rays to go beyond that limit. This result
indicates that it could be possible to see some effects about the
border of a black hole like light splitting depending on wave
length, in the same way of a rainbow like situation. Furthermore,
some effects are expectable for collisions with supermassive objects
due to the different penetration radius for radial orbits.
Altogether, these features can be considered as a skin effect. On
the other hand, non radial orbits aren't perturbed and no effect
seems to appear.



\end{document}